\documentclass[published]{JHEP3}

\JHEP{03(2002)023}

\newcommand{\beq}{\begin{equation}}
\newcommand{\eeq}{\end{equation}}

\title{Trialogue on the number of fundamental constants}

\author{Michael~J.~Duff\\
Michigan Center for Theoretical Physics, Randall Laboratory\\
Ann Arbor, MI, USA\\
E-mail: \email{mduff@umich.edu}}

\author{Lev  B. Okun\\
ITEP\\
Moscow, 117218, Russia\\

E-mail: \email{okun@heron.itep.ru}}

\author{Gabriele Veneziano\\
Theory Division, CERN\\
CH-1211 Geneva 23, Switzerland, and\\
Laboratoire de Physique Th\`eorique, Universit\`e Paris Sud\\
91405, Orsay, France\\
E-mail: \email{Gabriele.Veneziano@cern.ch}}

\abstract{This paper consists of three separate articles on the number
of fundamental dimensionful constants in physics.  We started our
debate in summer 1992 on the terrace of the famous CERN cafeteria.  In
the summer of 2001 we returned to the subject to find that our views
still diverged and decided to explain our current positions.  LBO
develops the traditional approach with three constants, GV argues in
favor of at most two (within superstring theory), while MJD advocates
zero.}

\accepted{March 9, 2002}
\keywords{Models of Quantum Gravity, Standard Model}
\received{March 4, 2002}

\begin{document}

\newpage
\part[Fundamental constants: parameters and units --- {\it L.B.~Okun}]
{\Large Fundamental constants: parameters and units}\label{partL}

\centerline{Lev B.  Okun}

\bigskip

\paragraph{Abstract.}

There are two kinds of fundamental constants of Nature: dimensionless
(like $\alpha\simeq1/137$) and dimensionful ($c$ --- velocity of light,
$\hbar$ --- quantum of action and angular momentum, and $G$ --- Newton's
gravitational constant). To clarify the discussion I suggest to refer
to the former as fundamental parameters and the latter as fundamental
(or basic) units. It is necessary and sufficient to have three basic
units in order to reproduce in an experimentally meaningful way the
dimensions of all physical quantities. Theoretical equations
describing the physical world deal with dimensionless quantities and
their solutions depend on dimensionless fundamental parameters. But
experiments, from which these theories are extracted and by which they
could be tested, involve measurements, i.e.\  comparisons with standard
dimensionful scales.  Without standard dimensionful units and hence
without certain conventions physics is unthinkable.

\section{Introduction: parameters and units}\label{secL1}

There is no well established terminology for the fundamental constants
of Nature.  It seems reasonable to consider as fundamental the
dimensionless ratios, such as the famous $\alpha=e^2/\hbar
c\simeq1/137$ and similar gauge and Yukawa couplings in the framework
of standard model of elementary particles or its extensions.

It is clear that the number of such constants depends on the
theoretical model at hand and hence depends on personal preferences
and it changes of course with the evolution of physics. At each stage
of this evolution it includes those constants which cannot be

expressed in terms of more fundamental ones, because of the absence of
the latter~\cite{1}. At present this number is a few dozens, if one
includes neutrino mixing angles. It blows up with the inclusion of
hypothetical new particles.

On the other hand the term ``fundamental constant'' is often used for
such dimensionful constants as the velocity of light $c$, the quantum
of action (and of angular momentum) $\hbar$, and the Newton
gravitational coupling constant $G$ . This article is concerned with
these dimensionful constants which I propose to call fundamental (or
basic) units.

Physics consists of measurements, formulas and ``words''. This article
contains no new formulas, it deals mainly with ``words'' because,
unlike many colleagues of mine, I believe that an adequate language is
crucial in physics. The absence of accurately defined terms or the
uses (i.e.\ actually misuses) of ill-defined terms lead to confusion
and proliferation of wrong statements.

\section{Stoney's and Planck's  units of L, T, M}\label{secL2}

The three basic physical dimensions: length L, time T, and mass M with
corresponding metric units: cm, sec, gram, are usually associated with
the name of C.F. Gauss. In spite of tremendous changes in physics,
three basic dimensions are still necessary and sufficient to express
the dimension of any physical quantity.  The number three corresponds
to the three basic entities (notions): space, time and matter. It does
not depend on the dimensionality of space, being the same in spaces of
any dimension. It does not depend on the number and nature of
fundamental interactions. For instance, in a world without gravity it
still would be three.

In the 1870's G.J. Stoney~\cite{2}, the physicist who coined the term
``electron'' and measured the value of elementary charge $e$,
introduced as universal units of Nature for L, T, M:
$l_S=e\sqrt{G}/c^2$, $t_S=e\sqrt{G}/c^3$, $m_S=e/\sqrt{G}$. The expression
for $m_S$ has been derived by equating the Coulomb and Newton
forces. The expressions for $l_S$ and $t_S$ has been derived from
$m_S$, $c$ and $e$ on dimensional grounds: $m_S c^2 = e^2/l_S$, $t_S =
l_S/c$.

When M. Planck discovered in 1899 $\hbar$, he introduced~\cite{3} as
universal units of Nature for L, T, M: $l_P=\hbar/m_P c$, $t_P=\hbar
/m_P c^2$, $m_P=\sqrt{\hbar c/G}$.

One can easily see that Stoney's and Planck's units are numerically
close to each other, their ratios being $\sqrt{\alpha}$.

\section{The physical meaning of units}\label{secL3}

The Gauss units were ``earth-bound'' and ``hand-crafted''. The cm and
sec are connected with the size and rotation of the
earth.\footnote{metre was defined in 1791 as a 1/40,000,000 part of
Paris meridian.} The gram is the mass of one cubic cm of water.

An important step forward was made in the middle of XX century, when
the standards of cm and sec were defined in terms of of wave-length
and frequency of a certain atomic line.

Enormously more universal and fundamental are $c$ and $\hbar$ given to
us by Nature herself as units of velocity $[v]=[L/T]$ and angular
momentum $[J]=[MvL]=[ML^2/T]$ or action $[S]=
[ET]=[Mv^2T]=[ML^2/T]$. (Here [ ] denotes dimension.)

\subsection{The meaning of $c$}

It is important that $c$ is not only the speed of light in
vacuum. What is much more significant is the fact that it is the
maximal velocity of any object in Nature, the photon being only one of
such objects. The fundamental character of $c$ would not be diminished
in a world without photons. The fact that $c$ is the maximal $v$ leads
to new phenomena, unknown in newtonian physics and described by
relativity. Therefore Nature herself suggests $c$ as fundamental unit
of velocity.

In the Introduction we defined as fundamental those constants
which cannot be calculated at our present level of fundamental
knowledge (or rather ignorance). This ``negative'' definition
applies equally to parameters and to units (to $\alpha$ and to
$c$).  At first sight $\alpha$ looks superior to $c$ because the
value of $\alpha$ does not depend on the choice  of units, whereas
the numerical value of $c$ depends explicitly on the units of
length and  time and hence on conventions. However $c$ is more
fundamental than $\alpha$ because its fundamental character has
not only a ``negative'' definition, but also a ``positive'' one:
it is the basis of relativity theory which unifies space and time,
as well as energy, momentum and mass.

By expressing $v$ in units of $c$ (usually it is defined as
$\beta=v/c$) one simplifies relativistic kinematics. On the other
hand the role of $c$ as a conversion factor between time and
distance or between mass and rest-energy is often overstated in
the literature. Note that in spite of the possibility of
measuring, say, distance in light-seconds, the length does not

become identical to time, just as momentum is not identical to
energy. This comes from the pseudoeuclidian nature of
four-dimensional space-time.

\subsection{The meaning of $\hbar$}

Analogously to $c$, the quantity $\hbar$ is is also fundamental in
the ``positive'' sense: it is the quantum of the angular momentum
$J$ and a natural unit of the action $S$. When $J$ or $S$ are
close to $\hbar$, the whole realm of quantum mechanical phenomena
appears.

Particles with integer $J$ (bosons) tend to be in the same state
(i.e.\ photons in a laser, or Rubidium atoms in a drop of
Bose-Einstein condensate). Particles with half-integer $J$
(fermions) obey the Pauli exclusion principle which is so basic
for the structure of atoms, atomic nuclei and neutron stars.

Symmetry between fermions and bosons, dubbed supersymmetry or
SUSY, is badly broken at low energies, but many theorists believe
that it is restored near the Planck mass (in particular in
superstrings and M-theories).

The role of  $\hbar$ as a conversion factor between frequency and
energy or between wave-length and momentum is often overstated.

It is natural when dealing with quantum mechanical problems to use
$\hbar$ as the unit of $J$ and $S$.

\subsection{The status of $G$}

The status of $G$ and its derivatives, $m_P$, $l_P$, $t_P$, is at
present different from that of $c$ and $\hbar$, because the quantum
theory of gravity is still under construction. The majority of experts
connect their hopes with extra spatial dimensions and
superstrings.\footnote{The characteristic length of a superstring
$\lambda_{s}$=$l_P/\sqrt{\alpha_{\rm GUT}}$, where ${\alpha_{\rm
GUT}}={\alpha(q^2={M^2}_{\rm GUT}})$. (As is well known, the
fundamental parameters are ``running'': their values depend on
$q^2$.)} But the bridge between superstrings and experimental physics
exists at present only as wishful thinking. Recent surge of interest
to possible modifications of Newton's potential at sub-millimetre
distances demonstrates that the position of $G$ is not as firm as that
of $c$ and $\hbar$.

\section{The cube of theories}\label{secL4}

The epistemological role of $c$, $\hbar$, $G$ units in classifying
theories was first demonstrated in a jocular article by G.~Gamov,
D.~Ivanenko and L.~Landau~\cite{g1}, then quite seriously by
M.~Bronshtein~\cite{b1,b2}, A.~Zelmanov~\cite{z1,z2} and others (see
e.g.~\cite{gg,o1}); and it is known now as the cube of theories.

The cube is located along three orthogonal axes marked by $c$
(actually by $1/c$), $\hbar$, $G$. The vertex ($000$) corresponds to
non-relativistic mechanics, ($c00$) --- to special relativity,
($0\hbar0$) --- to non-relativistic quantum mechanics, ($c\hbar0$) ---
to quantum field theory, ($c$0$G$) --- to general relativity,
($c$$\hbar$$G$) --- to futuristic quantum gravity and the Theory of
Everything, TOE. There is a hope that in the framework of TOE the
values of dimensionless fundamental parameters will be ultimately
calculated.

\section{The art of putting $c=1$, $\hbar=1$, $G=1$}\label{secL5}

The universal character of $c, \hbar,G$ and hence of $m_P, l_P, t_P$
makes natural their use in dealing with futuristic TOE. (In the case
of strings the role of $l_P$ is played by the string length
$\lambda_s$.) In such natural units all physical quantities and

variables become dimensionless. In practice the use of these units is
realized by putting $c=1$, $\hbar=1$, $G$ (or $\lambda_s$) $=1$ in all
formulas. However one should not take these equalities too literally,
because their left-hand sides are dimensionful, while the right-hand
sides are dimensionless. It would be more proper to use arrows
``$\rightarrow$'' (which mean ``substituted by'') instead of equality
signs ``$=$''.

The absence of $c,\hbar,G$ (or any of them) in the so obtained
dimensionless equations does not diminish the fundamental character of
these units. Moreover it stresses their universality and importance.

It is necessary to keep in mind that when comparing the theoretical
predictions with experimental results one has anyway to restore
(``$\leftarrow$'') the three basic units $c, \hbar, G$ in equations
because all measurements involve standard scales.

The above arguments imply what is often dubbed as a ``moderate
reductionism'', which in this case means that all physical phenomena

can be explained in terms of a few fundamental interactions of
fundamental particles and thus expressed in terms of three basic units
and a certain number of fundamental dimensionless parameters.

\section{International system  of units}\label{secL6}

An approach different from the above underlies the International
System of Units (Syst\'eme Internationale d'Unit\'{e}es ---
SI)~\cite{si,rp}. This System includes 7 basic units (metre, second,
kilogram, ampere, kelvin, mole, candela) and 17 derivative ones. The
SI might be useful from the point of view of technology and metrology,
but from the point of view of pure physics four out of its seven basic
units are evidently derivative ones. Electric current is number of
moving electrons per second. Temperature is up to a conversion factor

(Boltzman constant $k=1.38\times 10^{-23}$\,joules/kelvin) is the
average energy of an ensemble of particles. Mole is trivially
connected with the number of molecules in one gram-molecule, called
Avogadro's number $N_A=6.02\times 10^{23}$/mole. As for unit of
optical brightness or illumination (candela), it is obviously
expressed in terms of the flux of photons.

\looseness=1It is interesting to compare the character of $k$ with that of $c,
\hbar, m_P$. The Boltzman constant is an important conversion
factor which signals the transition from a few (or one) particle
systems to many particle systems. However it radically differs
from $c, \hbar, m_P$, as there is no physical quantity with the
dimension of $k$, for which $k$ is a critical value. The role of
conversion factor is only a secondary one for $c, \hbar, m_P$,
whereas for $k$ it is the only~one.

In the framework of SI vacuum is endowed with electric permittivity
$\varepsilon_0=8.85\times 10^{-12}$ farad/m and magnetic permeability
$\mu_0=12.57\times 10^{-17}$ newton/$(\mbox{ampere})^2$, whereas
$\varepsilon_0 \mu_0=1/c^2$.  This is caused by electrodynamic
definition of charge, which in SI is secondary with respect to the
current. In electrostatic units $\varepsilon_0=\mu_0=1$. According to
the SI standard this definition is allowed to use in scientific
literature, but not in text-books (see critical exposition of SI in
ref.~\cite{o2}).

\section[Remarks on Gabriele's part~II]{Remarks on Gabriele's part~\protect\ref{partG}}\label{secL7}

I note with satisfaction that some of the original arguments and
statements do not appear in his part of this Trialogue~\ref{partG}.
Among them there are the following statements: 1. that in string
theory there is room only for two and not three dimensionful
constants~\cite{15,16}; 2. that units of action are arbitrary [which
means that $\hbar$ is not a fundamental unit (LO)]; 3. that masses
unlike length and time intervals are not measured
directly~\cite{17}. Gabriele admits in section~\ref{secG6} that his
two units can be ``pedagogically confusing'' and the set $c, \hbar,
\lambda_{s}$ is ``most practical'', but he considers the latter ``not
economical'' and in other parts of the part~\ref{partG} he insists on
using ${\lambda_s}^{2}$ instead of $\hbar$.

Of course, if you forget about the pedagogical and practical sides of
physics, the most economical way is not to have fundamental units at
all, like Mike, but that is a purely theoretical approach (``{\tt
hep-th}''), and not physical one (``{\tt physics}'', ``{\tt
hep-ph}'').

It seems to me inconsistent to keep two units ($c, \lambda_s$)
explicitly in the equations, while substituting by unity the third one
($\hbar$), as Gabriele is doing in part~\ref{partG} and
refs.~\cite{15,16,17}. According to my section~\ref{secL5} above, this
corresponds to using $\hbar$ as a unit of $J$ and $S$, while not using
$c$ and $\lambda_s$ as units of velocity and length.

I also cannot agree that the electron mass, or $G_{F}$ are as good for
the role of fundamental unit as the Planck mass or $G$.

\section[Remarks on Mike's part~III]%
{Remarks on Mike's part~\protect\ref{partM}}\label{secL8}

In section~\ref{secM4} of Mike's part~\ref{partM} he introduces a
definition of fundamental constants with the help of an alien with
whom it is possible to exchange only dimensionless numbers. According
to Mike, only those constants are fundamental the values of which can
be communicated to the alien.  Thus Mike concludes that there exist no
fundamental units.  According to my section~\ref{secL5} above, this
actually corresponds to the use of $c, \hbar, G$ as fundamental units.

In fact, at the end of section~\ref{secM2} Mike writes ``that the most
economical choice is to use natural units where there are no
conversion factors at all.'' Mike explained to me that his natural
units are $c=\hbar=G=1$. As these equalities cannot be considered
literally, I believe that Mike uses the same three units as I
do. However he concludes section~\ref{secM2} with a statement:
``Consequently, none of these units or conversion factors is
fundamental.''

(In response to the above paragraph Mike added a new paragraph to his
section~\ref{secM2}, in which he ascribed to me the view that one
cannot put $c=1$. According to my section~\ref{secL5}, one can (and
should!) put $c=1$ in relativistic eqations, but must understand that
this means that $c$ is chosen as the unit of velocity.)

The ``alien definition'' of fundamental constants is misleading.
We, theorists, communicate not with aliens, but with our
experimental colleagues, students, and non-physicists. Such
communication is impossible and physics is unthinkable without
standardized dimensionful units, without conventions..

Concerning Mike's criticism of my article~\cite{o1}, I would like to
make the following remark. The statement that only dimensionless
variables, functions and constants have physical meaning in a theory
does not mean that every problem should be explicitly presented in
dimensionless form. Sometimes one can use dimensionful units and
compare their ratios with ratios of other dimensionful units. This
approach was used in ref.~\cite{o1}, where entertaining stories by
O.~Volberg~\cite{19} and G.~Gamov~\cite{20} were critically
analyzed. In these stories, in order to demonstrate the peculiarities
of relativistic kinematics, the velocity of light was assumed to be of
the order of that of a car, or even bicycle, while the everyday life
remained the same as ours. In ref.~\cite{o1} I have shown that if $c$
is changed, while dimensions of atoms are not changed (mass and charge
of electron as well as $\hbar$, are the same as in our world), then
electromagnetic and optical properties of atoms (and hence the
everyday life) would change drastically because of change of $\alpha$,
which is the ratio of electron velocity in hydrogen atom to that of
light. It is not clear to me why in section~\ref{secM5} of his paper
Mike disagrees with these considerations.

\section{Conclusions}\label{secL9}

It is obvious that using proper language (terms and semantics) three
fundamental units are the only possible basis for a selfconsistent
description of fundamental physics. Other conclusions are viable only
through the improper usage of terms.

\acknowledgments

I am very grateful to Gabriele and Mike for their patience in
discussing their points of view. I am indebted to Val Telegdi who
critically read the draft of this article and helped to improve it.
The hospitality of the CERN Theory Division and Michigan Center of
Theoretical Physics during the completion of this article is
appreciated, as well as Humboldt Award and the grant of RFBR
00-15-96562.

\newpage
\setcounter{section}{0}

\part[Fundamental units in physics: how many, if any? --- {\it
    G. Veneziano}]%
{\Large Fundamental units in physics: how many, if any?}\label{partG}

\centerline{Gabriele Veneziano}

\bigskip

\paragraph{Abstract.}

I summarize my previous work on the question of how many fundamental
dimensionful constants (fundamental units) are needed in various
theoretical frameworks such as renormalizable QFT + GR, old-fashioned
string theory, and modern string/M-theory. I will also try to
underline where past and present disagreement on these issues between
Lev Okun, Mike Duff, and myself appears to be originating from.

\section{Introductory remarks}\label{secG1}

Some fifteen years ago I wrote a short letter~\cite{15} on the number
of (dimensionful) fundamental constants in string theory, where I came
to the somewhat surprising conclusion that \emph{two} constants, with
dimensions of space and time, were both necessary and sufficient.
Somewhat later, I became aware of S.~Weinberg's 1983 paper~\cite{1},
whose way of looking at the question of defining fundamental constants
in physics I tried to incorporate in my subsequent work on the
subject~\cite{16,17}.

After reading those papers of mine once more, I still subscribe to
their content, even if I might have expressed some specific points
differently these days.  Here, rather than repeating the details of my
arguments, I will try to organize and summarize them stressing where,
in my opinion, the disagreement between Lev, Mike and myself arises
from. I have the impression that, in the end, the disagreement is more
in the words than in the physics, but this is what we should try to
find out.

The rest of this note is organized as follows: In section~\ref{secG2}
I make some trivial introductory statements that are hopefully

uncontroversial. In sections~\ref{secG3},~\ref{secG4} and~\ref{secG5}
I describe how I see the emergence of fundamental units (the name I
will adopt for fundamental dimensionful constants following Lev's
suggestion) in QFT+GR, in the old Nambu-Goto formulation of quantum
string theory (QST), and in the so-called Polyakov formulation,
respectively.  In sections~\ref{secG6} I will try to point at the
origin of disagreement between myself and Lev while, in
section~\ref{secG7}, the same will be done w.r.t.\ Mike.
Section~\ref{secG8} briefly discusses the issue of time-varying
fundamental units.

\section{Three questions and one answer}\label{secG2}

Let me start with two statements on which we all seem to agree:

\begin{itemize}
\item Physics is always dealing, in the end, with dimensionless
quantities, typically representing ratios of quantities having the
same dimensions, e.g.
\beq
\alpha = {e^2 \over \hbar c } \,,\qquad
{m_e \over m_p} \,, \qquad
\dots
\eeq

\item It is customary to introduce ``units", i.e.\ to consider the
ratio of any physical quantity $q$ to a fixed quantity $u_q$ \emph{of
the same kind} so that
\beq
q = (q/u_q) u_q \; ,
\eeq
where $u_q$ is a \emph{name} (e.g.\ \emph{centimetre} or
\emph{second}) and $(q/u_q)$ is a \emph{number}.  Obviously, $q_1/q_2 =
(q_1/u_q)/(q_2/u_q)$.

\item
Let us now ask the following three questions
\begin{itemize}
\item[$Q_1$:] are units arbitrary?

\item[$Q_2$:] are there units that are more fundamental than others
 according to  S.~Weinberg's definition~\cite {1}?

\item[$Q_3$:] How many units (fundamental or not) are necessary?
\end{itemize}

and try to answer them in the context of different theories of
elementary particles and interactions.

\end{itemize}

I hope we agree that the answer to the first question is yes, since
only $q_i/q_j$ matter and these ratios do not depend on the choice of
units.

I think that the answer to the other two questions depends on the
framework we are considering (Cf.\ Weinberg, ref.~\cite{1}).  The
next three sections therefore analyze $Q_2$ and $Q_3$ within three
distinct frameworks, and provide, for each case, answers $A_2$ and
$A_3$, respectively.

\section{Fundamental units in QFT+GR}\label{secG3}

Quantum Field Theory (QFT) (or more specifically the Standard Model
(SM)) plus General Relativity (GR) represent the state of the art in
HEP before the string revolution of 1984. Weinberg's 1983
paper~\cite{1} reflects therefore the attitude about FC's at the
dawn of the string revolution. I would summarize it briefly as
follows:
\begin{itemize}
\item  $A_2$: a qualified yes.

At the QFT level of understanding $c$ and $\hbar$ appear to be more
fundamental units of speed and action than any other. In newtonian
mechanics only the ratios of various velocities in a given problem
matter.  By contrast, in (special) relativity the ratio of each
velocity appearing in the problem to the (universal) speed of light,
$c$, also matters.  Likewise, in classical mechanics only the ratios
of various terms in the action matter, the overall normalization being
irrelevant while, in QM, the ratio of the total action to the
(universal) quantum of action $\hbar$ does matter (large ratios, for
instance, correspond to a semiclassical situation).  It appears
therefore that both $c$ and $\hbar$ have a special status as the most
basic units of speed and action.

Indeed, let's apply S.~Weinberg's criterion~\cite{1} and ask: can we
compute $c$ and $\hbar$ in terms of more fundamental units? Within QFT
the answer appears to be an obvious no.  Had we chosen instead some
other arbitrary units of speed and action, then, within a given
theory, we would be able to compute them, in principle at least, in
terms of $c$ and $\hbar$, i.e.\ in terms of something more fundamental
(and of some specified dimensionless constants such as $\alpha$).

\item $A_3$: most probably three

It is quite clear, I think, that in QFT+GR we cannot compute
everything that is observable in terms of $c$, $\hbar$, and of
dimensionless constants, without also introducing some mass or length
scale.  Hence it looks that the answer to the third question is indeed
three.  Unlike in the case of $c$ and $\hbar$, it is much less

obvious, however, which mass or length scale, if any, is more
fundamental in the sense of SW. The Planck mass, $M_P$, does not look
like a particularly good choice since it is very hard, even
conceptually, to compute, say, $m_e$ or $m_p$ in terms of $M_P$ in the
SM + GR framework.  This is a bit strange: we seem to need three
units, but we can only identify two fundamental ones. So why three?
Why not more?  Why not less?

Why not more? This is because it looks unnecessary (and even``silly"
according to present understanding of physical phenomena) to introduce
a separate unit for temperature, for electric current and resistance,
etc., or separate units for distances in the x, y and z directions. I
refer to Lev for a discussion about how to go from the seven units of
the International System of Units (SI) down to three~\cite{o2}, and
for how three fundamental units define the so-called ```cube" of
physical theories~\cite{o1}.

And why not less, say just two? Well because mass or energy appear as
concepts that are qualitatively different from, say, distances or time
intervals.  Let us recall how mass emerges in classical mechanics
(CM).  We can base CM on the action principle and get $F= ma$ by
varying the action
\beq
S = \int \left({1\over 2} \, m \dot{x}^2 - V(x) \right) dt
\Rightarrow m a = F \equiv - \frac{dV}{dx} \,,
\eeq
but, as it's well known, classically the action can be rescaled by an
arbitrary factor. If we had only one species of particles in Nature we
could use, instead of $S$,
\beq
\tilde{S} = \int \left({1\over 2}\, \dot{x}^2 - \frac{V(x)}{m} \right) dt
\equiv  \int \left({1\over 2} \, \dot{x}^2 - \tilde{V} (x)\right) dt
\Rightarrow  a = \tilde{F} \equiv - {1 \over m} \frac{dV}{dx}\,.
\eeq

No physical prediction would change by using units in which masses are
pure numbers provided we redefine forces accordingly!  In this system
of units $\hbar$ would be replaced by $\hbar/m$ and would have
dimensions of $v^2 \times t$. If we have already decided for $c$ as
unit of velocity, $\hbar$ would define therefore a fundamental unit of
time (the Compton wavelength of the chosen particle divided by $c$).
However, in the presence of many particles of different mass, we
cannot decide which mass to divide the action by, which choice is most
fundamental.

I think there is even a deeper reason why QFT+GR needs a separate unit
for mass.  QFT is affected by UV divergences that need to be
renormalized. This forces us to introduce a cut-off which, in
principle, has nothing to do with $c$, $\hbar$ or $M_P$, and has to be
``removed" in the end.  However, remnants of the cut-off remain in the
renormalized theory.  In QCD, for instance, the hadronic mass scale
(say $m_p$) originates from a mechanism known as dimensional
transmutation, and is arbitrary. Perhaps one day, through string

theory or some other unified theory of all interactions, we will
understand how $m_p$ is related to $M_P$, but in QFT+GR it is not. We
do not know therefore which of the two, $M_P$ or $m_p$, is more
fundamental and the same is true for the electron mass $m_e$, for
$G_F$ etc.\ etc.

The best we can do, in QFT+GR, is to take any one of these mass scales
(be it a particle mass or a mass extracted from the strength of a
force) as unit of mass and consider the ratio of any other physical
mass to the chosen unit as a pure number that, in general, we have no
way to compute, even in principle.
\end{itemize}

\section{Fundamental units in old-fashioned quantum string theory (QST)}\label{secG4}

\begin{itemize}
\item $A_2$: yes, $c$ and $\lambda_s$!

With string theory the situation changes because it is as if there
were a single particle, hence a single mass. Indeed, a single

classical parameter, the string tension $T$, appears in the Nambu-Goto
(NG) action:
\beq
S =  T \int d(\mbox{Area})\,, \qquad
\frac{S}{\hbar} = \lambda_s^{-2} \int d(\mbox{Area})\,,
\eeq
where the speed of light $c$ has already been implicitly used in order
to talk about the area of a surface embedded in space-time.  This fact
allows us to replace $\hbar$ by a well defined length, $\lambda_s$,
which turns out to be fundamental both in an intuitive sense and in
the sense of S.~Weinberg. Indeed, we should be able, in principle, to
compute any observable in terms of $c$ and $\lambda_s$ (see below for
an example).  Of course, I could instead compute $c$ and $\lambda_s$
in terms of two other physical quantities defining more down-to-earth
units of space and time, but this would not satisfy SW's criterion of
having computed $c$ and $\lambda_s$ in terms of something more
fundamental!

\item $A_3$: the above two constants are also sufficient!

This was the conclusion of my 1986 paper: string theory only needs two
fundamental dimensionful constants $c$ and $\lambda_s$, i.e.\ one
fundamental unit of speed and one of length.

The apparent puzzle is clear: where has our loved $\hbar$ disappeared?
 My answer was then (and still is): it changed its dress! Having
 adopted new units of energy (energy being replaced by energy divided
 by tension, i.e.\ by length), the units of action (hence of $\hbar$)
 have also changed.  And what about my reasoning in QFT+GR? Obviously
 it does not hold water any more: For one, QFT and GR get unified in
 string theory. Furthermore, the absence of UV divergences makes it
 unnecessary to introduce by hand a cut off.

And indeed the most amazing outcome of this reasoning is that the new
Planck constant, $\lambda_s^2$, \emph{is} the UV cutoff. We can
express this by saying that, in string theory, first quantization
provides the UV cutoff needed in order to make second quantization
well defined. Furthermore, in quantum string theory (QST), there are
definite hopes to be able to compute both $M_P$ and $m_p$ (in the
above string units, i.e.\ as lengths) in terms of $\lambda_s$, $c$ and
of a dimensionless parameter, the string coupling (see below).

The situation here reminds me of that of pure quantum gravity.  As
noticed by Novikov and Zeldovich~\cite[part~V, ch.~23,
par.~19]{NZ}, such a theory would only
contain \emph{two} fundamental units, $c$, and the Planck length $l_P
= \sqrt{G_N \hbar c^{-3}}$, but not $\hbar$ and $G_N$ separately.  We
may view string theory as an extension of GR that allows the
introduction of all elementary particles and all fundamental forces in
a geometrical way.  No wonder then to find that only geometrical units
are necessary.

Let us consider for instance, within the string theory framework, the
gravitational acceleration $a_2$ induced by a string of length $L_1$
on a string of length $L_2$ sitting at a distance $r$ from it. A
simple calculation gives (for $r \gg L_1, L_2$):
\beq
a_2 = g_s^2 c^2 \left(\frac{L_1}{r^2}\right),
\eeq
where $g_s$ is the (dimensionless!) string coupling discussed in the
next section. Clearly, the answer does not contain anything else but
geometrical quantities and a pure number.

Another more familiar example is the computation of the energy levels of atoms in terms of the electron mass, its charge, and $\hbar$.
These are given, to lowest order in $\alpha$, by \beq E_n = - {1 \over 2 n^2} m_e \left(\frac{e^2}{\hbar}\right)^2  =
 - {1 \over 2 n^2} (m_e c^2) \alpha^2
\eeq
Weinberg argues, convincingly I think, that the quantities $E_n$ are

less fundamental than the electron charge, mass and $\hbar$.  However,
if we argue that what we are really measuring are not energies by
themselves, but the transition frequencies
\beq
\omega_{mn} = {1 \over \hbar} (E_m - E_n) = {1 \over2} \left({1 \over
n^2} - {1 \over m^2}\right) {\alpha^2 c \over \lambda_s} \epsilon_e \,,
\eeq
we see that, once more, only $c$ and $\lambda_s$, and some in
principle calculable dimensionless ratios (such as the electron mass
in string units, $\epsilon_e = m_e/M_s$), appear in the
answer~\cite{15}.  Obviously, if we follow Weinberg's definition,
$\lambda_s$ and $\lambda_s/c$, and \emph{not} for instance $c
/\omega_{12}$ and $1/\omega_{12}$ (which are like the ``modern" units
of length, and time), play the role of fundamental units of length and
time.

\end{itemize}

\section{Fundamental units in modern QST/M-theory}\label{secG5}

We now turn to the same set of questions within the context of
first-quantized string theory in the presence of background
fields. Here I will attempt to give $A_2$ and $A_3$ together.  The
beautiful feature of this formulation is that all possible parameters
of string theory, dimensionful and dimensionless alike, are replaced
by background fields whose vacuum expectation values (VEV) we hope to
be able to determine dynamically. As a prototype, consider the bosonic
string in a gravi-dilaton background. The dimensionless action (i.e.\
the action divided by $\hbar$ in more conventional notation) reads:
\beq
\label{PA}
S = {1 \over 2} \int \sqrt{-\gamma}\left(\gamma^{\alpha\beta}
\partial_{\alpha} X^{\mu} {\partial}_{\beta}
 X^{\nu} G_{\mu\nu}(X) + R(\gamma) \phi(X)\right) d^2 z
\eeq
where $X^{\mu} = X^{\mu}(\sigma, \tau)$, $\mu = 0, 1 ,\dots, D-1$, are the
string coordinates as functions of the world-sheet coordinates $z
=(\sigma, \tau)$, with respect to which the the partial derivatives
are defined. Furthermore, $G_{\mu\nu}$ is the so-called string metric
and $\phi$ is the so-called dilaton. Finally, $\gamma_{\alpha\beta}$
and $R(\gamma)$ are, respectively, the metric and scalar curvature of
the two-dimensional Riemann surface having coordinates $\sigma$ and
$\tau$.  $\phi$ is clearly dimensionless, while the dimensions of the
metric components $G_{\mu\nu}$ are such that $G_{\mu\nu} X^{\mu}
X^{\nu}$ is also dimensionless.

The exponential of the expectation value of $\phi$ gives the
dimensionless parameter --- known as the string coupling $g_s$ ---
that controls the strength of all interactions (e.g.\ $\alpha$) and
thus also the string-loop expansion. Instead, the expectation value of
$G_{\mu\nu}$ converts lengths and time intervals into pure

numbers. Thus, through its non trivial dimension, the \emph{metric}
$G_{\mu\nu}$ actually provides the \emph{metre/clock}, i.e.\ the

fundamental units of space and time that we are after.

If the VEV of $G_{\mu\nu}$ is proportional to $\eta_{\mu\nu}$, the
flat minkowskian metric, then it will automatically introduce the
constants $c$ and $\lambda_s$ of the previous section via:
\beq
\label{GVEV}
\langle G_{\mu\nu}(X)\rangle = \mathop{\rm diag} \left( -c^2 \lambda_s^{-2},
\lambda_s^{-2}, \dots \right)
\eeq

The mere finiteness of $c$ and $\lambda_s$ is clearly of fundamental
importance.  However, in our context, the real question is: do the
actual values of $c$ and $\lambda_s$ mean something (in the same way
in which the actual value of $\langle \phi\rangle $ does)?  What is,
in other words, the difference between dimensionful and dimensionless
constants?  The answer is a bit subtle.  String theory should allow to
compute $\alpha$ in terms of the VEV of $\phi$. Similarly, it should
allow to compute $ (\Delta X)^2 \equiv G_{\mu\nu} \Delta X^{\mu}\Delta
X^{\nu}$ for some physical length $\Delta X^{\mu}$ (say for the
Hydrogen atom).  Calling that pure number so many centimetres would
fix the string length parameter in cm but, of course, this would be
just a convention: the truly convention-independent (physical)
quantity is just $(\Delta X)^2$.  Both $\langle \phi\rangle $ and
$(\Delta X)^2$ are pure numbers whose possible values distinguish one
theory (or one vacuum) from another.

The difference between the two kinds of constants, if any, simply
stems from the fact that, while different values of $\langle
\phi\rangle $ (or $\alpha$) define genuinely different theories,
values of $\langle G_{\mu\nu}\rangle $ that are related by a General
Coordinate Transformation (GCT) can be compensated by a GCT on $X$ and
thus define the same theory as long as $(\Delta X)^2$ remains the
same.  In particular, if $\langle G_{\mu\nu}\rangle \sim
\eta_{\mu\nu}$ as in the example discussed above, the actual
proportionality constants $c$ and $\lambda_s$ appearing in
(\ref{GVEV}) can be reabsorbed by a GCT.  This is why it does not make
sense to talk about the \emph{absolute} values of $c$ and $\lambda_s$
or to compare them to those of an alien: only the dimensionless
numbers $(\Delta X)^2$, i.e.\ the values of some physical length or
speed in those units are physically relevant and can be compared (see
section~\ref{secG7}).

The situation would be very different if $\langle G_{\mu\nu}\rangle $
would not be reducible to $\eta_{\mu\nu}$ via a GCT. That would mean a
really different world, like one with a different value of $\alpha$.
In ref.~\cite{GV4} I gave the example of $\langle G_{\mu\nu}\rangle $
proportional to the de-Sitter metric, stressing the fact that, in such
a vacuum, even $\lambda_s$ disappears in favour of a dimensionless
parameter similar to $\langle \phi\rangle $.  Thus, as stressed
in~\cite{16,17}, my early statement in~\cite{15} about having just two
constants should be considered valid if the vacuum of QST is
minkowskian, in particular in the NG formulation of QST.

To summarize, QM provides, through the string metric $G_{\mu\nu}$, a
truly fundamental metre/clock allowing us to reduce space-time
distances to pure numbers whose absolute value \emph{is} physically
meaningful.  Note, incidentally, that in Classical GR only $g_{\mu\nu}
\Delta X^{\mu}\Delta X^{\nu}$ is an invariant.  However, in the
classical case (and even for classical strings), only ratios of
quantities of this type matter while in QST, $(\Delta X)^2$ is, for
each single $\Delta X$, a meaningful pure number.

In conclusion, I still stand by my remark in~\cite{16} that the
fundamental constants of Nature are, in QST, the constants of the
vacuum.  How many (physically distinct) choices of its VEV's does QST
allow?  We now believe that all known consistent string theories
correspond to perturbations around different vacua of a single, yet
unknown, ``M-theory".  We still do not know, however, how many
physically inequivalent non-perturbative vacua M-theory has.  Until
then, I do not think we can really answer the question of fundamental
units in QST, but I would be very surprised if, in any consistent
string vacuum, we would find that we need more than one unit of length
and one of time.

\section{The disagreement with Lev}\label{secG6}

Lev cannot accept (part~\ref{partL}) that $\hbar$ has disappeared from
the list.  He claims that, without $\hbar$, there is no unit of
momentum, of energy, and, especially, of angular momentum.  But, as I
said in the previous two sections, $\hbar$ has not really disappeared:
it has actually been promoted, in string theory, to a grander role,
that of providing also, through QM, an UV cutoff that hopefully
removes both the infinities of QFT and ordinary Quantum Gravity and
the ubiquitous singularities of Classical GR.

I would concede, however, that, given the fact that momentum and
energy are logically distinct from lengths and times for ordinary
objects, insisting on the use of the same (or of reciprocal) units for
both sets can be pedagogically confusing. Therefore I do agree that
the set $c$, $\hbar$, and $\lambda_s$ define at present, within QST,
the most practical (though not the most economical) set of fundamental
units.

To rephrase myself: within the NG action there seems to be no reason
to introduce a tension $T$ or $\hbar$. The action is naturally
\emph{the} area and the Planck constant is the unit of area needed to
convert the action into a number.  However, by the standard definition
of canonically conjugate variables, this would lead to identical
dimensions for momenta and lengths (or for times and energies).  For
strings that's fine, since we can identify the energy of a string with
its length, but when it comes to ordinary objects, i.e.\ to
complicated bound states of fundamental strings or branes, it looks
less confusing to give momentum a unit other than length.  In order to
do that we introduce, somewhat artificially, a conversion factor, the
string tension $T$, so that energies are now measured in ergs, in GeV,
or whatever we wish, different choices being related by irrelevant
redefinitions of $T$.

\section{The disagreement with Mike}\label{secG7}

Two issues appear to separate Mike's position  from my own:
\begin{itemize}
\item The alien story

Mike quotes an example, due to Feynman, on how we could possibly tell
an alien to distinguish left from right. Then he asks: can we
similarly communicate to an alien our values for $c$ and $\lambda_s$
and check whether they agree with ours? I claim the answer to be: yes,
we can, and, to the same extent that the alien will be able to tell us
whether her\footnote{To stress that my alien's reaction is different
from that of Mike's alien I have also changed the alien's gender.}
$\alpha$ agrees with ours, she will also be able to tell us whether
her $c$ and $\lambda_s$ agree with ours.

In order to do that, we ``simply" have to give the alien \emph{our}
definitions of cm. and s.  in terms of a physical system she can
possibly identify (say the H atom) and ask: which are your values of
$c$ and $\lambda_s$ in these units?  If the alien cannot even identify
the system then she lives in a different world/string-vacuum; if she
does, then she should come up with the same numbers (e.g.\ $c= 3
\times 10^{10}$\,cm/s) or else, again, her world is not like ours. It
thus looks to me that the alien story supports the idea that we do
have, in our own world, some fundamental units of length and time.
Mike seems to agree with me on the alien's reply, but then concludes
that $c$ is \emph{not} a fundamental unit because a completely
rescaled world, in which both $c$ and the velocity of the electron in
the H atom are twice as large, is indistinguishable from ours.  I
conclude, instead, that $c$ \emph{is} a fundamental unit because the

velocity of our electron in units of $c$ \emph{is} a relevant number
to be compared with the alien's.

Incidentally, the same argument can be applied either to some
ancestors (or descendants) of ours, or to inequivalent string vacua. A
value of $c$ in cm/s for any of those which differs from ours would
really mean different worlds, e.g.\ worlds with different ratios of
the velocity of the electron in the Hydrogen atom and the speed of
light. We may \emph{either} express this by saying that, in the two
different worlds, $c$ is different in atomic units, \emph{or} by
saying that $c$ is the same but atomic properties differ. No
experimental result will be able to distinguish about these two
physically equivalent statements since a rescaling of all velocities
is inessential.

\item Reducing fundamental units to conversion factors

{\looseness=1Mike's second point is that these units can be used as conversion
factors, like $k_B$, in order to convert any quantity into any other
and, eventually, everything into pure numbers.  However, I do insist
that the point is \emph{not} to convert degrees Kelvin into MeV,
centimetres into seconds, or everything into numbers. The important
point is that there are units that are arbitrary and units that are
fundamental in the sense that, when a quantity becomes $O(1)$ in the

latter units, dramatic new phenomena occur.  It makes a huge
difference, for instance, having or not having a fundamental length.
Without a fundamental length, properties of physical systems would be
invariant under an overall rescaling of their size, atoms would not
have a characteristic size, and we would be unable to tell the alien
which atom to use as a metre.  By contrast, with a fundamental quantum
unit of length, we can meaningfully talk about short or large
distances (as compared to the fundamental length, of course).

}

Going back to the discussion at the end of section~\ref{secG5}, the
pure number $(\Delta X)^2$ has a meaning in itself.  In the absence of
any fundamental units of length and time I would be able to rescale
this number arbitrarily (e.g.\ by rescaling $G_{\mu\nu}$) without
changing physics.  Only ratios of two lengths in the problem, like
$(\Delta X_1)^2 /(\Delta X_2)^2$ would matter.  Because of QM,
however, there is a fundamental rod (and clock) that gives, out of any
\emph{single} physical length or time interval, \emph{a relevant} pure
number.

On this particular point, therefore, I tend to agree with Lev.  There
is, in relativity, a fundamental unit of speed (its maximal value);
there is, in QM, a fundamental unit of action (a minimal uncertainty);
there is, in string theory, a fundamental unit of length (the
characteristic size of strings). QST appears to provide the missing
third fundamental unit of the three-constants system.  These three
units form a very convenient system except that, classically, the
units of action are completely arbitrary (and the same is true
therefore of mass, energy etc.), while, quantum mechanically, only
$S/\hbar$ matters. In string theory this allows us to identify the
Planck constant with the string length eliminating the necessity, but
perhaps not the convenience, of a third unit besides those needed to
measure lengths and time intervals.

{\looseness=1I also agree with Mike that all that matters are pure numbers.  As I
stressed in section~\ref{secG2}, it is easy to convert any quantity
into a pure number by choosing arbitrarily some unit.  I only add to
this the observation that relativity and quantum mechanics provide, in
string theory, units of length and time which look, at present, more
fundamental than any other.  The number of distinct physical
quantities (and of corresponding units) is a matter of choice and
convenience, and also depends on our understanding of the underlying
physical laws.  Within QFT + GR it looks most useful to reduce this
number to three, but there is no obvious candidate for the third unit
after $c$ and $\hbar$. With QST, the third unit naturally emerges as
being the string length $\lambda_s$. However there appears an
interesting option to do away with $\hbar$. Going further down, say
from two to one or to zero, means considering space as being
\emph{equivalent} to time or as both being \emph{equivalent} to pure
numbers, while, keeping the two units $c$ and $\lambda_s$, allows to
\emph{express} space and time intervals \emph{in terms} of pure
numbers.}

This is what distinguishes, in my opinion, fundamental units from
conversion factors. While I see no reason to distinguish the units of
temperature from those of energy, and thus to introduce Boltzmann's
constant, I see every reason to distinguish space from time and to
introduce $c$ as a fundamental unit of speed and \emph{not} as a
trivial conversion factor. Another clear difference is that, while the
ratio $E(T)/T$ is always the same, we do observe, in Nature, a variety
of speeds (all less than $c$, so far), of lengths, and of frequencies.

\end{itemize}

\section{Time variation of fundamental units?}\label{secG8}

I think that the above discussion clearly indicates that the ``time
variation of a fundamental unit", like $c$, has no meaning, unless we
specify what else, having the same units, is kept fixed.  Only the
time variation of dimensionless constants, such as $\alpha$ or
$(\Delta X)^2$ for an atom have an intrinsic physical meaning.

We do believe, for instance, that in a cosmological background the
variation in time of $G_{\mu\nu}$ is accompanied by a corresponding
variation of the $\Delta X^{\mu}$ of an atom so that $(\Delta X)^2$
remains constant.  The same is usually assumed to be true for
$\alpha$. However, this is not at all an absolute theoretical
necessity (e.g.\ $\alpha$ can depend on time, in QST, if $\phi$ does),
and should be (and indeed is being) tested.  For instance, the same
$(\Delta X)^2$ is believed to grow with the expansion of the Universe
if $\Delta X^{\mu}$ represents the wavelength of light coming to us
from a distant galaxy.  The observed red shift only checks the
\emph{relative} time-dependence of $(\Delta X)^2$ for an atom and for
the light coming from the galaxy.

However, I claim that, in principle, the time variation of $(\Delta
X)^2$ has a physical meaning for each one of the two systems
separately because it represents the time variation of some physical
length w.r.t. the fundamental unit provided by string theory. For
instance, in the early Universe, this quantity for the CMBR photons
was much smaller than it is today ($O(10^{30})$). If it ever
approached values $O(1)$, this may have left an imprint of
short-distance physics on the CMBR spectrum.

\acknowledgments

I am grateful to Lev and Mike for having given their serious thoughts
to this issue and for pushing me to clarify my point of view to them,
and to myself.  I also wish to acknowledge the support of a ``Chaire
Internationale Blaise Pascal", administered by the ``Fondation de
L'Ecole Normale Sup\'erieure'', during the final stages of this work.

\newpage

\setcounter{section}{0}

\part[A party political broadcast on behalf of the Zero Constants\\
Party --- {\it M.J.~Duff}]%
{\Large A party political broadcast on behalf of the Zero Constants
Party}\label{partM}

\centerline{Michael J.~Duff}

\bigskip

\paragraph{Abstract.}

According to the manifesto of Okun's Three Constants Party, there are
three fundamental dimensionful constants in Nature: Planck's constant,
${\hbar}$, the velocity of light, $c$, and Newton's constant, $G$.
According to Veneziano's Two Constants Party, there are only two: the
string length $\lambda_{2}^'$ and $c$.  Here we present the
platform of the Zero Constants Party.

\section{The false propaganda of the Three Constants Party}\label{secM1}

As a young student of physics in high school, I was taught that there
were three basic quantities in Nature: Length, Mass and
Time~\cite{Feather}.  All other quantities, such as electric charge or
temperature, occupied a lesser status since they could all be
re-expressed in terms of these basic three.  As a result, there were
three basic units: centimetres, grams and seconds, reflected in the
three-letter name ``CGS'' system (or perhaps metres, kilograms and
seconds in the alternative, but still three-letter, ``MKS'' system).

Later, as an undergraduate student, I learned quantum mechanics,
special relativity and newtonian gravity. In quantum mechanics, there
was a minimum quantum of action given by Planck's constant ${\hbar}$;
in special relativity there was a maximum velocity given by the
velocity of light $c$; in classical gravity the strength of the force
between two objects was determined by Newton's constant of gravitation
$G$. In terms of length, mass, and time their dimensions are
\begin{eqnarray}
[c]&=&LT^{-1}
\nonumber\\{}
[\hbar]&=&L^{2}MT^{-1}
\nonumber\\{}
[G]&=&L^{3}M^{-1}T^{-2}\,.
\end{eqnarray}
Once again, the number three seemed important and other dimensionful
constants, such as the charge of the electron $e$ or Boltzmann's
constant $k$, were somehow accorded a less fundamental role.  This
fitted in perfectly with my high school prejudices and it seemed
entirely natural, therefore, to be told that these three dimensionful
constants determined three basic units, first identified a century ago
by Max Planck, namely the Planck length $L_{P}$, the Planck mass
$M_{P}$ and the Planck time $T_{P}$:
\begin{eqnarray}
L_P&=&\sqrt{G\hbar/c^3}=1.616\times10^{-35}\,{\rm m}
\nonumber\\
M_P&=&\sqrt{\hbar c/G}=2.177\times10^{-8}\,{\rm kg}
\nonumber\\
T_P&=&\sqrt{G\hbar/c^5}=5.390\times10^{-44}\,{\rm s}
\label{Planck}
\end{eqnarray}

Yet later, researching into quantum gravity which attempts to combine

quantum mechanics, relativity and gravitation into a coherent unified
framework, I learned about the Bronshtein-Zelmanov-Okun (BZO)
cube~\cite{o1}, with axes $\hbar$, $c^{-1}$ and $G$, which neatly
summarizes how classical mechanics, non-relativistic quantum
mechanics, newtonian gravity and relativistic quantum field theory can
be regarded respectively as the $({\hbar},c^{-1},G) \rightarrow 0$,
$(c^{-1},G) \rightarrow 0$, $({\hbar},c^{-1}) \rightarrow 0$, and $(G)
\rightarrow 0$ limits of the full quantum gravity.  Note, once again
that we are dealing with a three-dimensional cube rather than a square
or some figure of a different dimension.

What about Kaluza-Klein theories which allow for
$D>4$ spacetime dimensions?  Unlike $\hbar$ and $c$, the dimensions of
$G$ depend on $D$:
\begin{equation}
[G_{D}]=M^{-1}L^{D-1}T^{-2}
\end{equation}
and hence (dropping the $P$ subscript), the $D$-dimensional Planck
length $L_{D}$, mass $M_{D}$ and time $T_{D}$ are given by
\begin{eqnarray}
L_D{}^{D-2}&=&G_{D}\hbar c^{-3}
\nonumber\\
M_D{}^{D-2}&=&G_{D}{}^{-1}\hbar^{D-3}c^{5-D}
\nonumber\\
T_D{}^{D-2}&=&G_{D}\hbar c^{-1-D}\,.
\label{PlanckD}
\end{eqnarray}
After compactification to four dimensions, $G\equiv G_{4}$ then
appears as
\begin{equation}
\frac{1}{G_{4}}=\frac{1}{G_{D}}V\,,
\end{equation}
where $V$ is the volume of the compactifying manifold. Since $V$ has
the four-dimensional interpretation as the vacuum expectation value of
scalar modulus fields coming from the internal components of the
metric tensor, it depends on the choice of vacuum but does not
introduce any more fundamental constants into the lagrangian.

Adherents of this conventional view of the fundamental constants of
Nature have been dubbed the ``Three Constants Party'' by Gabriele
Veneziano~\cite{17}.  Lev Okun is their leader.  Until
recently I was myself, I must confess, a card-carrying
member.\footnote{It seems that the choice of length, mass and time as
the three basic units is due to Gauss~\cite{NIST}, so we could declare
him to be the founder of the Three Constants Party, although this was
long before the significance of $c$ and $\hbar$ was appreciated.}

\section{The false propaganda of the Two Constants Party}\label{secM2}

My faith in the dogma was shaken, however, by papers by Gabriele
\cite{15,16,17}, self-styled leader of the
rebel Two Constants Party.  As a string theorist, Gabriele begins with
the two-dimensional Nambu-Goto action of a string.  He notes that,
apart from the velocity of light still needed to convert the time
coordinate $t$ to a length coordinate $x^{0}=ct$, the action divided
by $\hbar$ requires only one dimensionful parameter, the string length
$\lambda_{2}^'$ (denoted $\lambda_{s}$ by Gabriele).
\begin{equation}
\lambda_{2}{}^{2}=\frac{\hbar}{cT_{2}}\,,
\end{equation}
where $T_{2}=1/2\pi c\alpha'$ is the tension of the string and
$\alpha'$ is the Regge slope. This is because the Nambu-Goto action
takes the form
\begin{equation}
\frac{S_{2}}{\hbar}=\frac{1}{\lambda_{2}{}^{2}}{\rm Area}
\label{string}
\end{equation}
So if this were to describe the theory of everything (TOE), then the
TOE would require only two fundamental dimensionful constants $c$ and
$\lambda_{2}$. In superstring theory, the ten-dimensional Planck
length is given in terms of the string length $\lambda_{2}^'$ and
the vacuum expectation value of the dilaton field $\phi$
\begin{equation}
L_{10}{}^{2}=\lambda_{2}{}^{2}\langle e^{\phi}\rangle
\end{equation}
Once again, the vev of $\phi$ will be different in different vacua but
does not introduce any new constants into the lagrangian.

A similar argument for reducing the three constants $h,c,G$ to just
two was made previously by Zeldovich and Novikov~\cite{NZ} with regard
to quantum gravity. The Einstein-Hilbert action divided by $\hbar$
involves $G$ and $\hbar$ only in the combination $G\hbar$ appearing in
the square of the Planck length, and so we need only $L_{P}$ and
$c$. Of course quantum gravity does not pretend to be the TOE and so
this argument still leaves open the number of dimensionful constants
required for a TOE.

In the light of the 1995 M-theory~\cite{Duff} revolution, we might
wish to update Gabriele's argument by starting with the corresponding
three-dimensional action for the M2-brane,
\begin{equation}
\frac{S_{3}}{\hbar}=\frac{1}{\lambda_{3}{}^{3}}(\hbox{3d-volume})\,,
\label{membrane}
\end{equation}
where the corresponding parameter is the membrane length
$\lambda_{3}$.
\begin{equation}
\lambda_{3}{}^{3}=\hbar/cT_{3}
\end{equation}
and where $T_{3}$ is the membrane tension. Alternatively, we could
start with the six-di\-men\-sio\-nal action of the dual M5-brane,
\begin{equation}
\frac{S_{6}}{\hbar}=\frac{1}{\lambda_{6}{}^{6}}(\hbox{6d-volume})
\label{fivebrane}
\end{equation}
where the corresponding parameter is the fivebrane length $\lambda_{6}$
\begin{equation}
\lambda_{6}{}^{6}=\frac{\hbar}{cT_{6}}
\end{equation}
and where $T_{6}$ is the fivebrane tension. Eleven-dimensional
M-theory is, in fact, simpler than ten-dimensional superstring theory
in this respect, since there is no dilaton.  Consequently, the three
lengths: membrane length, fivebrane length and eleven-dimensional
Planck length are all equal~\cite{DLM} up to calculable numerical
factors $\lambda_{3}\sim\lambda_{6}\sim L_{11}^'$.  So the
fundamental length in M-theory is $\lambda_{3}$ rather than
$\lambda_{2}$ and will be shorter for string coupling less than
unity~\cite{Shenker}.

However, even if we substitute $\lambda_{3}$ for $\lambda_{2}$,
Gabriele would say that we are still left with the number two.  This
also reduces the number of basic units to just two: length and time.

Gabriele's claim led to many heated discussions in the CERN cafeteria
between Lev, Gabriele and myself. We went round and round in
circles. Back at Texas A\&M, I continued these arguments at lunchtime
conversations with Chris Pope and others.  There at the College
Station Hilton, we eventually reached a consensus and joined what
Gabriele would call the Zero Constants Party~\cite{17}.

Our attitude was basically that $\hbar$, $c$ and $G$ are nothing but
conversion factors e.g.\ mass to length, in the formula for the
Schwarzschild radius $R_{S}$
\[
R_{S}=\frac{2Gm}{c^{2}}\,,
\]
or energy to frequency
\[
E=\hbar \omega
\]
energy to mass
\[
E=mc^{2}
\]
no different from Boltzmann's constant, say, which relates energy to
temperature
\[
E=kT\,.
\]
As such, you may have as many so-called ``fundamental'' constants
as you like; the more different units you employ, the more
different constants you need.\footnote{In this respect, I take the
the number of dimensionful fundamental constants to be synonymous
with the number of fundamental (or basic) units.}  Indeed, no
less an authority than the \emph{Conf\'{e}rence G\'{e}n\'{e}rale
des Poids et Mesures}, the international body that administers the
SI system of units, adheres to what might be called the Seven
Constants Party, decreeing that seven units are ``basic'':
metre(length), kilogram (mass), second (time), ampere (electric
current), kelvin (thermodynamic temperature), mole (amount of
substance), candela (luminous intensity), while the rest are
``derived''~\cite{NIST,SI}.  The attitude of the Zero Constants
Party is that the most economical choice is to use natural units
where there are no conversion factors at all.  Consequently, none
of these units or conversion factors is fundamental.

Incidentally, Lev (part~\ref{partL}) objects in his
section~\ref{secL5} that equations such as $c=1$ cannot be taken
literally because $c$ has dimensions.  In my view, this apparent

contradiction arises from trying to use two different sets of units at
the same time, and really goes to the heart of my disagreement with
Lev about what is real physics and what is mere convention.  In the
units favored by members of the Three Constants Party, length and time
have different dimensions and you cannot, therefore, put $c=1$ (just
as you cannot put $k=1$, if you want to follow the conventions of the
Seven Constants Party).  If you want to put $c=1$, you must trade in
your membership card for that of (or at least adopt the habits of) the
Two Constants Party, whose favorite units do not distinguish length
from time.\footnote{This $(\hbar,G)$ wing of the Two Constants Party
is different from Gabriele's $(c,\lambda_{2})$ wing, which prefers not
to introduce a separate unit for mass.}  In these units, $c$ \emph{is}
dimensionless and you may quite literally set it equal to one.  In the
natural units favored by the Zero Constants Party, there are no
dimensions at all and $\hbar=c=G=\cdots=1$ may be imposed literally
and without contradiction.  With this understanding, I will still
refer to constants which have dimensions in some units, such as
$\hbar,c,G,k\ldots$, as ``dimensionful constants'' so as to
distinguish them from constants such as $\alpha$, which are
dimensionless in any units.

\section{Three fundamental theories?}\label{secM3}\label{three}

Lev and Gabriele remain unshaken in their beliefs, however.  Lev
(part~\ref{partL}) makes the, at first sight reasonable, point (echoed
by Gabriele in part~\ref{partG}) that $\hbar$ is more than just a
conversion factor.  It embodies a fundamental physical principle of
quantum mechanics that there is a minimum non-zero angular momentum.
Similarly, $c$ embodies a fundamental physical principle of special
relativity that there is a maximum velocity $c$.  If I could
paraphrase Lev's point of view it might be to say that there are three
``fundamental'' units because there are three fundamental physical
theories: quantum mechanics, special relativity and gravity.
According to this point of view, temperature, for example, should not
be included as a basic unit (or, equivalently, Boltzmann's constant
should not be included as a fundamental constant.)

However, I think this elevation of $\hbar$, $c$ and $G$ to a special
status is misleading.  For example, the appearance of $c$ in
$x^{0}=ct$ is for the benefit of people for whom treating time as a
fourth dimension is unfamiliar.  But once you have accepted $O(3,1)$
as a symmetry the conversion factor becomes irrelevant.  We have
become so used to accepting $O(3)$ as a symmetry that we would not dream
of using different units for the three space coordinates,\footnote{I
am grateful to Chris Pope for this example.} but to be perverse we
could do so.

To drive this point home, and inspired by the \emph{Conf\'{e}rence
G\'{e}n\'{e}rale des Poids et Mesures}, let us introduce three new
superfluous units: xylophones, yachts and zebras to measure intervals
along the $x$, $y$ and $z$ axes.  This requires the introduction of
three superfluous ``fundamental'' constants, $c_{x}$, $c_{y}$ and
$c_{z}$ with dimensions length/xylophone, length/yacht and
length/zebra, respectively, so that the line element becomes:
\begin{equation}
ds^{2}=-c^{2}dt^{2}+c_{x}{}^{2}dx^{2}+c_{y}{}^{2}dy^{2}+c_{z}{}^{2}dz^{2}\,.
\end{equation}
Lev's point is that the finiteness of $c$ ensures that we have
$O(3,1)$ symmetry rather than merely $O(3)$.  This is certainly true.
But it is equally true that the finiteness of $c_{x}$, say, ensures
that we have $O(3,1)$ rather than merely $O(2,1)$.  In this respect,
the conversion factors $c$ and $c_{x}$ are on an equal
footing.\footnote{To put this more rigorously, the Poincar\'e group
admits a Wigner-In\"{o}n\"{u} contraction to the Galileo group,
obtained by taking the $c\rightarrow \infty$ limit.  However, this is
by no means unique.  There are other contractions to other subgroups.
For example, one is obtained by taking the $c_{x}^'\rightarrow
\infty$ limit.  Although of less historical importance, these other
subgroups are mathematically on the same footing as the Galileo group.
So, in my opinion, the singling out of $c$ for special treatment has
more to do with psychology than physics.}  Both are, in Gabriele's
terminology (part~\ref{partG}), equally ``silly''.  Both can be set
equal to unity and forgotten about.

Similarly, the ``fundamental'' lengths $\lambda_{d}^'$ appearing in
brane actions~(\ref{string}),~(\ref{membrane}) and (\ref{fivebrane})
can be removed from the equations by defining new dimensionless
worldvolume coordinates, $\xi'$, related to the old ones, $\xi$, by
$\xi=\lambda_{d}\xi '$.

So I would agree with Lev that the finiteness of the conversion
factors is important (minimum angular momentum, maximum velocity) but,
in my view, no significance should be attached to their value and you
can have as many or as few of them as you like.

The reason why we have so many different units, and hence conversion
factors, in the first place is that, historically, physicists used
different kinds of measuring apparatus: rods, scales, clocks,
thermometres, electroscopes etc.  Another way to ask what is the
mimimum number of basic units, therefore, is to ask what is, in
principle, the minimum number of basic pieces of apparatus.\footnote{I
am grateful to Chris Isham for this suggestion.}  Probably Lev,
Gabriele and I would agree that $E=kT$ means that we can dispense with
thermometers, that temperature is not a basic unit and that
Boltzmann's constant is not fundamental.  Let us agree with Lev that
we can whittle things down to length, mass and time or rods, scales
and clocks.  Can we go further?  Another way to argue that the
conversion factor $c$ should not be treated as fundamental, for
example, is to point out that once the finiteness of $c$ has been
accepted, we do not need both clocks and rulers.  Clocks alone are
sufficient since distances can be measured by the time it takes light
to travel that distance, $x=ct$.  We are, in effect, doing just that
when we measure interstellar distances in light-years.  Conversely, we
may do away with clocks in favor of rulers.  It is thus superfluous to
have both length and time as basic units.  Similarly, we can do away
with rulers as basic apparatus and length as a basic unit by trading
distances with masses using the formula for the Compton wavelength
$R_{C}=h/mc$.  Indeed, particle theorists typically express length,
mass and time units as inverse mass, mass and inverse mass,
respectively.  Finally, we can do away with scales by expressing
particle masses as dimensionless numbers, namely the ratio of a
particle mass to that of a black hole whose Compton wavelength equals
its Schwarzschild radius.  So in this sense, the black hole acts as
our rod, scale, clock, thermometer etc.\ all at the same time.  In
practice, the net result is as though we set $\hbar=c=G=\cdots=1$ but
we need not use that language.

J-M.~Levy-LeBlond~\cite{leblond} puts it like this: ``This, then, is
the ordinary fate of universal constants: to see their nature as
concept synthesizers be progressively incorporated into the implicit
common background of physical ideas, then to play a role of mere unit
conversion factors and often to be finally forgotten altogether by a
suitable redefinition of physical units.''

\section{An operational definition}\label{secM4}

{\it ``If, however, we imagine other worlds, with the same physical laws as
those of our own world, but with different numerical values for the
physical constants determining the limits of applicability of the old
concepts, the new and correct concepts of space, time and motion, at
which modern science arrives only after very long and elaborate
investigations, would become a matter of common knowledge.''}

\rightline{\small George Gamow, \emph{Mr.\  Tompkins in paperback}~\cite{20}}

\medskip

It seems to me that this issue of what is fundamental will continue to
go round and around until we can all agree on an operational
definition of ``fundamental constants''.  Weinberg~\cite{1}
defines constants to be fundamental if we cannot calculate their
values in terms of more fundamental constants, not just because the
calculation is too hard, but because we do not know of anything more
fundamental.  This definition is fine, but does not resolve the
dispute between Gabriele, Lev and me.  It is the purpose of this
section to propose one that does.  I will conclude that, according to
this definition, the dimensionless parameters, such as the fine
structure constant, are fundamental, whereas all dimensionful
constants, including $\hbar$, $c$ and $G$, are not.\footnote{My
apologies to those readers to whom this was already blindingly
obvious.  A similar point of view may be found in \cite{Gribbin}.  On
the other hand, I once read a letter in Physics World from a
respectable physicist who believed that a legitimate ambition of a TOE
would be to calculate the numerical value of $\hbar$.}

In physics, we frequently encounter ambiguities such as ``left or
right'' and ``matter or antimatter''.  Let us begin by recalling
Feynman's way of discriminating between what are genuine differences
and what are mere conventions. Feynman imagines that we can
communicate with some alien being~\cite{Feynman}.  If it were not for
the violation of parity in the weak interactions we would have no way
of deciding whether what he\footnote{I will follow Feynman and assume
that the alien is a ``he'', without resolving the ``he or she''
ambiguity.} calls right and left are the same as what we call right
and left.  However, we can ask him to perform a cobalt 60 experiment
and tell him that the spinning electrons determine a left handed
thread.  In this way we can agree on what is left and right.  When we
eventually meet the alien, of course, we should beware shaking hands
with him if he holds out his left hand (or tentacle).  He would be
made of antimatter and annihilate with us!  Fortunately, after the
discovery of CP violation we could also eliminate this ambiguity.

In a similar vein, let us ask whether there are any experiments that
can be performed which would tell us whether the alien's universe has
the same or different constants of nature as ours.  If the answer is
yes, we shall define these constants to be fundamental, otherwise not.
In particular, and inspired by Gamow's Mr.\ Tompkins~\cite{20}, we
will ask whether there is in principle any experimental difference
that would allow us to conclude unambiguously that his velocity of
light, his Planck's constant or his Newton's constant are different
from ours.  By ``unambiguously'' I mean that no perceived difference
could be explained away by a difference in conventions.  (Of course,
even Feynman's criterion is not devoid of theoretical assumptions.  We
have to assume that the cobalt behaves the same way for the alien as
for us etc.  To be concrete, we might imagine that we are both
described by a TOE (perhaps M-theory) in which the fundamental
constants are given by vacuum expectation values of scalar fields.
The alien and we thus share the same lagrangian but live in possibly
different vacua.  Let us further assume that both vacua respect
$O(3,1)$ symmetry.)

\section{The operationally indistinguishable world of Mr.\  Tompkins}\label{secM5}

The idea of imagining a universe with different constants is not new,
but, in my opinion, the early literature is very confusing.  For
example, Vol'berg~\cite{19} and Gamow~\cite{20} imagine a
universe in which the velocity of light is different from ours, say by
ten orders of magnitude, and describe all sorts of weird effects that
would result:

{\it ``The initials of Mr.\  Tompkins originated from three fundamental
physical constants: the velocity of light $c$; the gravitational
constant $G$; and the quantum constant $h$, which have to be changed
by immensely large factors in order to make their effect easily
noticeable by the man on the street.''}

\rightline{\small George Gamow, \emph{Mr.\  Tompkins in paperback}~\cite{20}}

\medskip

In this one sentence, Gamow manages to encapsulate everything I am
objecting to!  First, he takes it as axiomatic that there are three

fundamental constants.  Second, he assumes a change in these constants
can be operationally defined.  I for one am mystified by such

comparisons.  After all, an inhabitant of such a universe (let us
identify him with Feynman's alien) is perfectly free to choose units
in which $c=1$, just as we are.  To use the equation
\[
k=\frac{E}{c}
\]
to argue that in his universe, for the same energy $E$, the photon
emitted by an atom would have a momentum $k$ that is ten orders of
magnitude smaller than ours is, to my mind, meaningless. There is no
experimental information that we and the alien could exchange that
would allow us to draw any conclusion.

By contrast, in his critique of Vol'berg and Gamow, Lev~\cite{o1}
imagines a universe in which the binding energy of an electron in a
hydrogen atom $E=me^{4}/\hbar^{2}$ exceeds twice the electron rest
energy $2mc^{2}$, where $m$ and $e$ are the electron mass and charge
respectively.  In such a universe it would be energetically favorable
for the decay of the proton to a hydrogen atom and a positron $p
\rightarrow H+e^{+}$.  This universe is demonstrably different from
ours.  But, in my opinion, the correct conclusion has nothing to do
with the speed of light, but simply that in this universe the
dimensionless fine structure constant $\alpha=e^{2}/\hbar c$ exceeds
$\sqrt{2}$.

I believe that these two examples illustrate a general truth: no
experimental information that we and the alien could exchange can
unambiguously determine a difference in dimensionful quantities.  No
matter whether they are the $\hbar$, $c$ and $G$ sacred to the Three
Constants Party, the $\lambda_{2}$ and $c$ of the Two Constants Party
or the seven constants of the \emph{Conf\'{e}rence G\'{e}n\'{e}rale
des Poids et Mesures}.  Any perceived difference are all merely
differences in convention rather than substance.  By contrast,
differences in dimensionless parameters like the fine structure
constants are physically significant and meaningful.\footnote{In his
section~\ref{secG7}, Gabriele (part~\ref{partG}) claims to disagree
with me on this point, but I think the first two sentences of his
section~\ref{secG8} indicate that we are actually in agreement.  If,
for example, the alien tells us that he observes the decay $p
\rightarrow H+e^{+}$, then we can be sure that his $\alpha$ is
different from ours.  Choosing to attribute this effect (or any other
effect) to a difference in $c$ rather than $\hbar$ or $e$, however, is
entirely a matter of convention, just as the difference between left
and right would be a matter of convention in a world with no CP
violation.  So $c$ fails the Feynman test.}  Of course, our current
knowledge of the TOE is insufficient to tell us how many such
dimensionless constants Nature requires.  There are 19 in the Standard
model, but the aim of M-theory is to reduce this number.  Whether they
are all calculable or whether some are the result of cosmological
accidents (like the ratios of distances of planets to the sun) remains
one of the top unanswered questions in fundamental
physics.\footnote{Indeed, participants of the Strings 2000 conference
placed it in the top ten~\cite{Strings2000}.}

\section{What about theories with time-varying constants?}\label{secM6}

Suppose that our ``alien'' came not from a different universe but from a
different epoch in our own universe and we stumbled
across his historical records. In this way of thinking, the issue of
whether $\hbar$, $c$ and $G$ are fundamental devolves upon the issue of
whether the results of any experiments could require the unambiguous
conclusion that $\hbar$, $c$ and $G$ are changing in time. According
to our criterion above, any such time-dependence would be merely
convention, without physical significance.

On the other hand, many notable physicists, starting with
Dirac~\cite{Dirac}, have nevertheless entertained the notion that $G$
or $c$ are changing in time. (For some reason, time-varying $\hbar$ is
not as popular.) Indeed, papers on time-varying $c$ are currently in
vogue as as an alternative to inflation. I believe that these ideas,
while not necessarily wrong, are frequently presented in a misleading
way and that the time-variation in the physical laws is best described
in terms of time-varying dimensionless ratios, rather than
dimensionful constants.\footnote{This point of view is also taken
in~\cite{Barrow}.} So, in my opinion, one should talk about time
variations in the dimensionless parameters of the standard model but
not about time variations in $\hbar$, $c$ and $G$. For example, any
observed change in the strength of the gravitational force over
cosmological times should be attributed to changing mass ratios rather
than changing $G$.  For example, the proton is approximately
$10^{19}$ times lighter than the black hole discussed in
section~\ref{secM3}, whose Compton wavelength equals its Schwarzschild
radius.  It is then sensible to ask whether this dimensionless ratio
could change over time.\footnote{One could then sensibly discuss a
change in the number of protons required before a star reaches its
Chandrasekar limit for gravitational collapse.  I am grateful to Fred
Adams for this example.}

Unfortunately, this point was made insufficiently clear in the recent
paper presenting astrophysical data suggesting a time-varying fine
structure constant~\cite{Webb}. As a result, a front page article in
the New York Times~\cite{Glanz} announced that the speed of light
might be changing over cosmic history.\footnote{I am reminded of the
old lady who, when questioned by the TV interviewer on whether she
believed in global warming, responded: ``If you ask me, it's all this
changing from Fahrenheit to Centigrade that causing it!''.}

In the context of M-theory which starts out with no parameters at all,
these standard model parameters would appear as vacuum expectation
values of scalar fields.\footnote{The only other possibility
compatible with maximal four-dimensional spacetime symmetry is the
vacuum expectation value of a $4$-index field strength.  For example,
the cosmological constant can receive a contribution from the vev of
the M-theory $4$-form~\cite{DVN}.}  Indeed, replacing parameters by
scalar fields is the only sensible way I know to implement time
varying constants of Nature.  The role of scalar fields in determining
the fundamental constants in a TOE was also emphasized by
Gabriele~\cite{15,16,17}.

\section{Conclusions}\label{secM7}

The number and values of fundamental dimensionless constants appearing
in a Theory of Everything is a legitimate subject of physical enquiry.
By contrast, the number and values of dimensionful constants, such as

$h$, $c$, $G$,\ldots\ is a quite arbitrary human construct, differing
from one choice of units to the next.  There is nothing magic about
the choice of two, three or seven.  The most economical choice is
zero.  Consequently, none of these dimensionful constants is
fundamental.

\acknowledgments

My interest in this subject was fired by conversations with Lev Okun
and Gabriele Vene\-ziano.  I am also grateful for discussions on
fundamental constants with colleagues at Texas A\&M, especially Chris
Pope, at the University of Michigan, especially Gordy Kane, and at
Imperial College, especially Chris Isham.  I have enjoyed
correspondence with Thomas Dent, Harald Fritsch, James Glanz, Robert
C.  Helling, Saibal Mitra, Jose Sande Lemos, Warren Siegel and Steve
Weinberg.  Alejandro Rivero points out that the Greek root of
\emph{dia} in \emph{dialogue} means \emph{through} and not \emph{two}
so \emph{Trialogue} is misnamed.  The hospitality of the High Energy
Theory Group, Imperial College and the Mathematical Institute, Oxford,
during the preparation of this article is gratefully acknowledged.

This research was supported in part by DOE Grant DE-FG02-95ER40899.

\paragraph{Note added.}

Warren Siegel (private communication) makes the following interesting points:
\begin{enumerate}
\item Planck was actually a member of the Four Constants Party, since
his original paper introduced not only a basic length, mass and time
but also a temperature.\footnote{By analyzing Planck's papers~\cite{3}
Lev came to the conclusion that by adding $k$ to $c$, $h$ and $G$,
Planck contradicts his definition of natural units \cite{Okun2}.}

\item In 1983, the \emph{Conf\'{e}rence G\'{e}n\'{e}rale des Poids et
Mesures} declared $c$ to have the value 299,792,458 metres/second
exactly, \emph{by definition}, thus emphasizing its role as a nothing
but a conversion factor.\footnote{So asking whether the value of $c$
has changed over cosmic history is like asking whether the number of
litres to the gallon has changed.}

\item Sailors use the perverse units of section~\ref{secM3}, when they
measure intervals along the $x$ and $y$ axes in nautical miles and
intervals along the $z$ axis in fathoms.  The same observation was
made independently by Steve Weinberg (private communication).
\end{enumerate}


\newpage

\begin{thebibliography}{99}

\bibitem{1}
S. Weinberg, \emph{Overview of theoretical prospects for understanding
the values of fundamental constants}, in \emph{The constants of
physics}, W.H.  McCrea and M.J.  Rees eds., \emph{Phil.\  Trans.\ R.\
Soc.\  London} {\bf A310} (1983) 249.


\bibitem{2}
G.J. Stoney, \emph{The philosophical magazine and journal of science},
{\bf 11} (1881) 381.


\bibitem{3}
M. Planck, \emph{\"Uber irrevesible Strahlungsvorg\"{a}nge}, \emph{S.-B. Preuss
Akad.\ Wiss.\ }(1899) 440-480;
\emph{Ann.\ d.\ Phys.\ }{\bf 1} (1900) 69 reprinted in
\emph{Max Planck, Physikalische Abhandlungen und
Vortr\"{a}ge}, \emph{Band I.\ Friedr.\ Vieweg.\ }1958, pp.~560--600, pp.~614--667.


\bibitem{g1}
G. Gamov, D. Ivanenko and L. Landau, \emph{Zh.\ Russ.\ Fiz.\ Khim.\ Obstva.\
Chast' Fiz.\ }{\bf 60} (1928) 13 (in Russian).

\bibitem{b1}
M. Bronshtein, \emph{K voprosu o vozmozhnoy teorii mira kak tselogo, (On
a possible theory of the world as a whole)}, in \emph{Osnovnye
problemy kosmicheskoy fiziki (Basic problems of cosmic physics)}.
Kiev, ONTI (1934), pp.~186--218, in particular p.~210 (in Russian).

\bibitem{b2}
M. Bronshtein, \emph{Physikalische Zeitscrift der Sowjetunion} {\bf 9}
(1936) 140.

\bibitem{z2}
A. Zelmanov, \emph{Kosmologia (Cosmology)}, in \emph{Razvitie astronomii
v SSSR (Development of astronomy in USSR)}, Nauka, Moscow, 1967,
pp.~320--390, in particular p.~323 (in Russian).


\bibitem{z1}
A. Zelmanov, \emph{Mnogoobrazie materialnogo mira i problema
beskonechnosti Vselennoi (Diversity of the material world and the
problem of infinity of the Universe)}, in \emph{Beskonechnost' i
Vselennaya (Infinity and Universe)}, Mysl', Moscow, 1969, pp.~274--324, in particular p.~302 (in Russsian).

\bibitem{gg}
G. Gorelik, \emph{Razmernost' prostranstva (Dimension of space)},
MGU, Moscow, 1983, chapter~5 (in Russian).

\bibitem{o1}
L.B. Okun, \emph{The fundamental constants of physics},
\spu{34}{1991}{818}.

\bibitem{si}
\emph{Symbols, units and nomenclature in physics: document U.I.P.}
     {\bf 20} (1978), International Union of Pure and Applied Physics,
     S.U.N. Commission.

\bibitem{rp}
\emph{Review of Particle Properties}, \prd{45}{1992}{} Part II;
page III.4.


\bibitem{o2}
L.B. Okun, \emph{Particle physics: the quest for the substance of
substance}, Harwood academic publishers, Chur, New York 1985,
Appendix 1. On systems of physical units.


\bibitem{15}

G. Veneziano, \emph{A stringy nature needs just two constants},
\epl{2}{1986}{199}.

\bibitem{16}
G. Veneziano, \emph{Quantum strings and the constants of nature}, in
\emph{The challenging questions},  proceedings of
the 27th Course of International School held July 26 - August 3,
1989, Erice, Italy, A.~Zichichi ed., Plenum Press, New York 1990, p.~199.

\bibitem{17}
G. Veneziano, \emph{Fundamental constants in field and string theory},
CERN-TH.6725/92, talk given at the 6th session of the
International Workshop of Theoretical Physics, in \emph{String
quantum gravity and physics at the Planck scale}, Erice, Sicily,
21-28 June 1992, N. Sanchez and A. Zichichi eds.,
World Scientific 1993, p.~552.



\bibitem{19}
O.A. Volberg, \emph{Zanimatel'naya progulka v strane Einshteina (An entertaining trip into Einstein's country)}, in \emph{Zanimatel'naya
Mekhanika (Entertaining Mechanics)}, Ya.I.~Perel'man ed., GRN-PYuL, Leningrad - Moscow 1935, ch.~11, p.~188 (in Russian).

\bibitem{20}
G. Gamow, \emph{Mr.\ Tompkins in paperback}, containing \emph{Mr.\ Tompkins
in wonderland} and \emph{Mr.\ Tompkins explores the atom}, Cambridge
University Press, Cambridge, England 1988.


\bibitem{NZ}
Ya.B.~Zeldovich and I.D.~Novikov, \emph{The structure and evolution of
the universe}, University of Chicago press 1982.

\bibitem{GV4}
G. Veneziano, \emph{Physics with a fundamental length}, in
\emph{Physics and mathematics of strings,
 Vadim Knizhnik Memorial Volume},
 L.~Brink, D.~Friedan and A.M.~Polyakov eds.,
WSPC, 1990,  p.~509.



\bibitem{Feather}
N.~Feather,
\emph{Mass, length and time}, Edinburgh University Press, 1959.

\bibitem{Duff}
M.J.  Duff, \emph{The world in eleven dimensions: supergravity,
supermembranes and M-theory}, I.O.P.  Publishing 1999,
\href{http://bookmark.iop.org/bookpge.htm/book=815p}%
{\tt http://bookmark.iop.org/bookpge.htm/book=815p}.

\bibitem{DLM}
M.J. Duff, J.T. Liu and R.~Minasian,
\emph{Eleven-dimensional origin of string/string duality: a one loop
test}, \npb{452}{1995}{261}.

\bibitem{Strings2000}
M.J. Duff,
\emph{The top ten questions in fundamental physics}, in \emph{Strings 2000},
proceedings of the International superstrings conference, University
of Michigan, Ann Arbor, M.J.~Duff, J.T.~Liu and Lu eds., World Scientific, 2001,
\href{http://feynman.physics.lsa.umich.edu/strings2000/millennium.html}%
{\tt http://feynman.physics.lsa.umich.edu/strings2000/millennium.html}.


\bibitem{Dirac}
P.A.M. Dirac, \emph{The cosmological constants}, \emph{Nature} {\bf
139} (1937).


\bibitem{Shenker}
M.R.~Douglas, D.~Kabat, P.~Pouliot and S.H.~Shenker, \emph{D-branes
and short distances in string theory}, \npb{485}{1997}{85}
[\hepth{9608024}].

\bibitem{NIST}
The NIST reference on constants, units and uncertainty,
\href{http://physics.nist.gov/cuu/Units/international.html}%
{\tt http://physics.nist.gov/cuu/Units/international.html}.

\bibitem{SI}
Particle Data Group, \emph{Review of particle properties},
\prd{45}{1992}{S1},  part~II; page III.4., erratum \ibid{D46}{1992}{5210}.



\bibitem{leblond}
J.-M.  Levy-Leblond, \emph{The importance of being (a) constant}, in
\emph{Problems in the foundations of physics}, Enrico Fermi School
LXXII, G.~Torraldi ed., (North Holland 1979), p.~237.

\bibitem{Feynman}
R.P. Feynman,
\emph{The character of physical law}, Penguin, 1992, p.~101.

\bibitem{DVN}
M.J. Duff and P. van Nieuwenhuizen, \emph{Quantum inequivalence of
different field representations}, \plb{94}{1980}{179}.

\bibitem{Gribbin}
J. Gribbin and P. Wesson,
\emph{Fickle constants of physics}, \emph{New Scientist}, 4 July 1992, p.~30.

\bibitem{Barrow}
J.D.~Barrow and J.~Magueijo,
\emph{Varying-$\alpha $ theories and solutions to the cosmological
problems}, \astroph{9811072}.

\bibitem{Webb}
J.K.~Webb  et al.,
\emph{Further evidence for cosmological evolution of the fine structure
constant}, \astroph{0012539}.

\bibitem{Glanz}
J.~Glanz and D.~Overbye, \emph{Anything can change, it seems, even an
immutable law of nature}, \emph{New York Times}, August 15, 2001.


\bibitem{Okun2}
L.~Okun, \emph{Cube or hypercube of natural units?}, \hepph{0112339}.



\end{thebibliography}
\end{document}